\documentclass[reprint,amsmath,amssymb,aps,prc,superscriptaddress]{revtex4-2}
\usepackage{graphicx}
\usepackage{epstopdf}
\usepackage{dcolumn}
\usepackage{bm}
\usepackage{subcaption}
\usepackage[justification=Justified,font=smaller]{caption}
\usepackage{booktabs}
\usepackage{hyperref}
\usepackage{mathtools}
\usepackage{color}
\hypersetup{hypertex=true,
			colorlinks=true,
			linkcolor=blue,
			anchorcolor=blue,
			citecolor=blue,
			}
\begin{document}


\title{Impact of dark matter on strange quark stars described by different quark models}


\author{Yida Yang}
\affiliation{School of Physics and Electronic Science, East China Normal University, Shanghai 200241, China\\}
\author{Chen Wu}
\affiliation{Xingzhi College, Zhejiang Normal University, Jinhua, 321004, China.\\}
\author{Ji-Feng Yang}
\affiliation{School of Physics and Electronic Science, East China Normal University, Shanghai 200241, China\\}

\date{\today}

\begin{abstract}

Dark matter is hypothesized to interact with ordinary matter solely through gravity and may be present in compact objects such as strange quark stars. We treat strange quark stars admixed with dark matter as two-fluid systems to investigate the potential effects of dark matter on strange quark stars.
Quark matter is described by the quasiparticle model and the extended MIT bag model for comparison. Dark matter is treated as asymmetric, self-interacting, and composed of massive fermionic particles. The two-fluid Tolman-Oppenheimer-Volkoff (TOV) equations are employed to solve for specific stellar properties.
Our analysis yields relations between central energy density and mass, radius and mass, as well as tidal deformability and mass. The calculated curves generally align with observational data. In particular, we find  that the pattern in which fermionic asymmetric dark matter affects the properties of strange quark stars may not be influenced by the equation of state (EOS) of strange quark matter.
\end{abstract}

\maketitle
\section{Introduction}
If strange quark matter is the true ground state of strongly interacting matter \cite{PhysRevD.30.272,bodmerCollapsedNuclei1971,farhiStrangeMatter1984,yinSlowlyRotatingNeutron2010}, then there could exist a self-bound compact object composed of deconfined $u,~d, ~\mathrm{and}~s$ quarks \cite{weberStrangeQuarkMatter2005,buballaNJLmodelAnalysisDense2005}. Strange quark stars are such hypothetical compact objects. They can be formed by the conversion of neutron stars \cite{chuQUARKMATTERSYMMETRY2013,bombaciQuarkDeconfinementImplications2004,staffThreeStageModel2007}, but their macroscopic properties would differ from those of traditional neutron stars. At the same mass, strange quark stars tend to have smaller radii, and both their mass and radius can be arbitrarily small \cite{glendenningCompactStars2000}. Due to these characteristics, strange quark stars have become an attractive research subject in astrophysics and particle physics.

In addition to ordinary matter, there is evidence suggesting the existence of dark matter in the universe \cite{rubinRotationalProperties211980,rubinRotationAndromedaNebula1970,refregierWeakGravitationalLensing2003,Tyson_1998,Lewis_2003,Clowe_2006,bertoneHistoryDarkMatter2018,trimbleExistenceNatureDark1987}. It does not interact directly with ordinary matter, causing detection difficulties \cite{hongImpactsSymmetryEnergy2023,kouvarisConstrainingAsymmetricDark2011}. However, it produces significant gravitational effects, such as on compact objects like neutron stars. Dark matter may appear inside compact stars through capture processes \cite{kainDarkMatterAdmixed2021,roblesImprovedTreatmentDark2023,goldmanWeaklyInteractingMassive1989,bertoneCompactStarsDark2008,kouvarisCanNeutronStars2010} or accumulation during stellar formation \cite{deliyergiyevDarkCompactObjects2019,zentnerAsymmetricDarkMatter2011}. When dark matter admixes with ordinary matter in sufficient amounts, it will influence some observable properties of compact stars (such as mass-radius relation and tidal deformability) \cite{kainDarkMatterAdmixed2021,hongImpactsSymmetryEnergy2023}. There has been considerable research on how dark matter affects compact objects such as white dwarfs and neutron stars \cite{bramanteDarkMatterCompact2024}. Since strange quark stars are also a type of compact object, incorporating dark matter can expand our research on strange quark stars and in turn explore potential properties of dark matter \cite{hongMixedDarkMatter2024}.

In this work, we study strange quark stars admixed with dark matter as a two-fluid system \cite{hongMixedDarkMatter2024,hippertDarkMatterRegular2023,mukhopadhyayQuarkStarsAdmixed2016}. This requires solving the two-fluid Tolman-Oppenheimer-Volkov (TOV) equations \cite{sandinEffectsMirrorDark2009} by incorporating the equations of state for quarks and dark matter.
For strange quark matter, there are currently no reliable first-principle quantum chromodynamics (QCD) calculations \cite{baymHadronsQuarksNeutron2018,klahnModernCompactStar2007} of equation of state for strange quark matter at high densities. To this end, many phenomenological models have been developed, such as the MIT bag model \cite{mukhopadhyayQuarkStarsAdmixed2016,chodosNewExtendedModel1974}, Nambu–Jona-Lasinio (NJL) model \cite{buballaNJLmodelAnalysisDense2005,nambuDynamicalModelElementary1961,hatsudaQCDPhenomenologyBased1994,liStrangeQuarkStars2020}, quasiparticle model \cite{peshierEquationStateDeconfined2000,peshierEffectiveModelQuarkgluon1994,peshierMassiveQuasiparticleModel1996,zhaoEQUATIONSTATEQUASIPARTICLE2010,bannurSelfconsistentQuasiparticleModel2008,bannurSelfconsistentQuasiparticleModel2007}, and confined-density-dependent-mass (CDDM) model \cite{pengMassFormulasThermodynamic1999}. For comparison, we use the quasiparticle model \cite{peshierEffectiveModelQuarkgluon1994,liStrangeQuarkMass2021,liStructuresStrangeQuark2019} and the extended MIT bag model \cite{schertlerStrangeMatterEffective1997,schertlerQuarkPhasesNeutron2000,schertlerMediumEffectsStrange1997} to describe strange quark matter.

Due to the unclear origin and properties of dark matter, a rich variety of models have been proposed in literature. Examples include weakly interacting massive particles (WIMPs), sterile neutrinos, axions, mirror particles, supersymmetry particles, fermionic/bosonic asymmetric dark matter, etc (for a review, see \cite{bertoneHistoryDarkMatter2018,bertoneParticleDarkMatter2005,fengDarkMatterCandidates2010}). The development of these dark matter models are motivated by various reasons, such as to solve the gauge hierarchy problem or the strong CP problem. In this work, we consider the possibility of zero-temperature fermionic asymmetric dark matter with self-interactions \cite{mukhopadhyayQuarkStarsAdmixed2016,narainCompactStarsMade2006}. Here, the asymmetric nature of dark matter prevents its self-annihilation \cite{kainDarkMatterAdmixed2021}.

This paper is organized as follows. In Sec.~\ref{sec2}, the models  to describe strange quark matter are presented. In Sec.~\ref{sec3}, we show the model of dark matter considered in this work. The TOV equations for two-fluid systems and the corresponding equations for solving tidal deformability are introduced in Sec.~\ref{sec4}. Our numerical calculation results regarding the properties of strange quark stars admixed with dark matter are presented in Sec.~\ref{sec5}. Included in Sec.~\ref{sec6} our the summaries and discussions.
\section{Strange quark matter\label{sec2}}
Considering the electroweak reactions in quark stars, strange quark matter needs to fulfill the following chemical equilibrium conditions and electric charge neutrality condition:
\begin{eqnarray}
	&&
	\mu_{d}=\mu_{u}+\mu_{e}, \label{eq1} \\
	&&
	\mu_{s}=\mu_{u}+\mu_{e}, \label{eq2} \\
	&&
	\frac23\rho_{u} =\frac{1}{3}\rho_{s}+\frac{1}{3}\rho_{d}+\rho_{e}, \label{eq3}
\end{eqnarray}
which are necessary for calculating the energy density and pressure of strange quark matter. In this work, two different models are used to describe strange quark matter.

\subsection{Quasiparticle model}
The quasiparticle model can not only be applied to small clusters of strange particles or hadronic physics \cite{wuStrangeQuarkMatter2015}, but also to high-density strange quark matter \cite{peshierEquationStateDeconfined2000,zhaoEQUATIONSTATEQUASIPARTICLE2010}.
In the quasiparticle model of quark-gluon plasma (QGP) , the microscopic interaction between quarks and gluons is treated as an effective mass modification of each particle \cite{liStructuresStrangeQuark2019,liStrangeQuarkMass2021,liTidalDeformabilitiesRadii2021,peshierEquationStateDeconfined2000}. The system of interacting quarks and gluons can be effectively represented as an ideal gas of non-interacting quasiparticles with a temperature and density dependent mass \cite{zhaoEQUATIONSTATEQUASIPARTICLE2010,vanheugtenFermiliquidTheoryImbalanced2012}.


The effective quark mass that depends on chemical potential $\mu$ for massive quark reads \cite{peshierEquationStateDeconfined2000,bannurSelfconsistentQuasiparticleModel2008,peshierEffectiveModelQuarkgluon1994}:
\begin{eqnarray}
	m_f^2(\mu)=(m_{f0}+m_q(\mu))^2+m_q^2(\mu)\label{eq4},
\end{eqnarray}
with (zero temperature) \cite{braatenSimpleEffectiveLagrangian1992,vijaBraatenPisarskiMethodFinite1995,peshierEquationStateDeconfined2000}
\begin{eqnarray}
	m_q^2(\mu)=\frac{N_f\mu^2g^2(\mu)}{18\pi^2}\label{eq5},
\end{eqnarray}
where $N_f$ is the number of quark flavors, $m_{f0}$ is the mass of current quark $(u,~d,~\mathrm{or}~s)$. And $g^2(\mu)$ is the effective coupling constant ,which has the following relation with the two-loop approximate running coupling constant \cite{bannurSelfconsistentQuasiparticleModel2007,caswellAsymptoticBehaviorNonAbelian1974},
\begin{eqnarray}
	\alpha_s(\mu)=\frac{g^2(\mu)}{4\pi}=\frac{6\pi}{(33-2N_f)\mathrm{ln}(a\mu)}\\\nonumber
	\times\left[1-\frac{3(153-19N_f)}{(33-2N_f)^2}\frac{\ln(2\ln(a\mu))}{\ln(a\mu)}\right]\label{eq6},
\end{eqnarray}
where $a=1.91/(2.91\xi)$, and $\xi$ is a phenomenological parameter of the quasiparticle model.

The quark number density for each quark flavour reads \cite{liStrangeQuarkMass2021},
\begin{eqnarray}
	\rho_f(\mu)=\frac{N_c}{3\pi^2}\left(\mu^2-m_f^2(\mu)\right)^{3/2}\theta(\mu-m_f(\mu)).
\end{eqnarray}
There is a step function $\theta(\mu-m_f(\mu))$ on the right-hand side and chemical potential $\mu$ changes within a certain range.

Then, according to \cite{zongCalculationEquationState2008}, the pressure of strange quark matter at zero temperature and finite chemical potential can be written as,
\begin{eqnarray}
	P(\mu)=P(\mu)|_{\mu=0}+\sum_{i=u,d,s,e}\int_{0}^{\mu_i}\rho_i(\mu')\mathrm{d}\mu',
\end{eqnarray}
where $P(\mu)|_{\mu=0}$ is called the vacuum pressure which defines
the pressure at $\mu=0$, which can be analogized to the non-vanishing vacuum pressure in the MIT bag model \cite{buballaNJLmodelAnalysisDense2005}. Thus in the present model the vacuum pressure can be defined as $P(\mu)|_{\mu=0}\equiv-B(B>0)$, analogous to the bag constant in MIT bag model \cite{chodosNewExtendedModel1974}. In order to  preserves the confinement of quarks and gluons, the phenomenological parameter $B$ is always a positive to ensure negative vacuum pressure.

In strange quark stars, the active freedoms are $u,~d,~\mathrm{and}~s$ quarks and leptons ($e ~\mathrm{and}~ \mu$), so the pressure is given by, 
\begin{eqnarray}
	P(\mu)=-B+P_u(\mu_u)+P_d(\mu_d)+P_s(\mu_s)+P_e(\mu_e).
\end{eqnarray}

It is obvious that $\mu_d~\mathrm{and}~\mu_s$ can be expressed in terms of $\mu_u~\mathrm{and}~\mu_e$ according to Eq.~(\ref{eq1}) and Eq.~(\ref{eq2}). Therefore, by treating $\mu_u$ as a variable parameter and allowing it to vary within a certain range, the Eq.~(\ref{eq3}) can be used to solve for $\mu_e$. 

Finally, the energy density of this model is given by,

\begin{eqnarray}
	\begin{aligned}\epsilon&=-P+\mu\cdot\frac{\partial P}{\partial\mu}\\&=-P+\sum_{i=u,d,s,e}\mu_{i}\rho_{i}(\mu_{i}).\end{aligned}
\end{eqnarray}

One can see that the pressure and energy density in this quasiparticle model depend on the current quark masses, the parameter $\xi$ , and the vacuum pressure $B$. In this work, we adopt the following fixed parameter values $\xi=63~\mathrm{MeV}$, $B^{1/4}=129~\mathrm{MeV}$, $m_{u0}=m_{d0}=0~\mathrm{MeV}$ and treat $m_{s0}$ as a free parameter, as per Ref.~\cite{liStrangeQuarkMass2021}.

\begin{figure}
	\includegraphics[width=0.5\textwidth]{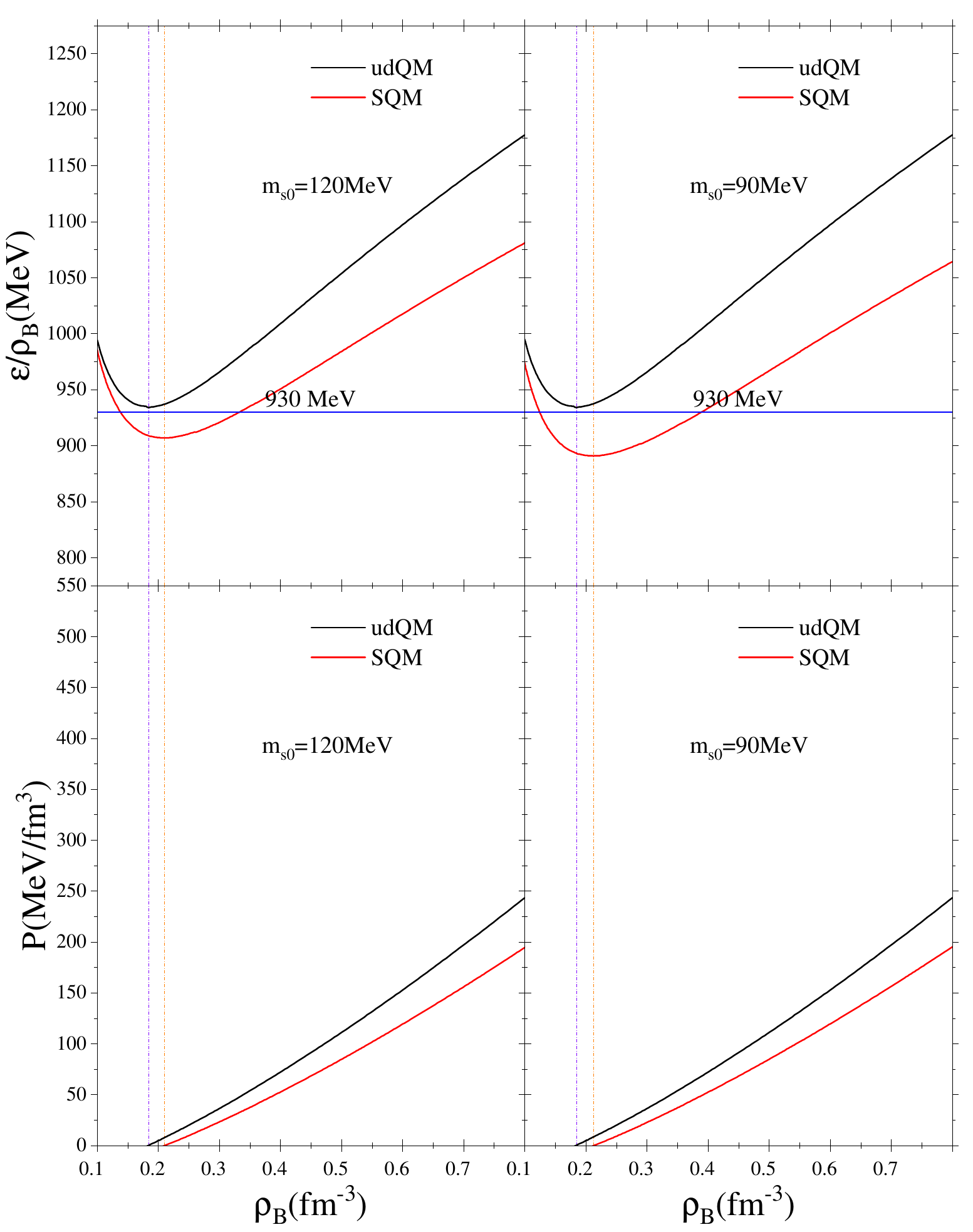}
	\caption{\label{qp}The energy per baryon $\varepsilon/\rho_B$ and pressure $P$ as functions of baryon number density ($\rho_B$) for SQM (red line) and udQM (black line) described by the quasiparticle model at zero temperature. The current strange quark mass $m_{s0}$ is 120 MeV for left panels and 90 MeV for right panels. }
\end{figure}

Before calculating the EOSs $P( \varepsilon )$, some verifications need to be performed. As previously mentioned, strange quark matter (SQM) may be the true ground state of strongly interacting matter. Therefore, the minimum energy per baryon of SQM should be less than 930 MeV at zero temperature, while the minimum energy per baryon of u-d quark matter (udQM) should be larger than 930 MeV (absolutely stable condition). This value of 930 MeV corresponds to the minimum energy per baryon observed stable nuclei $\prescript{56}{ }{Fe}$. Furthermore, the baryon number density ($\rho_B$) at zero pressure ($P$) should coincide with the baryon number density at which the energy per baryon reaches ($\varepsilon/\rho_B$) its minimum. To verify whether these conditions are satisfied, we calculate both $\varepsilon/\rho_B$ and $P$ as functions of $\rho_B$. The results are shown in Fig.~\ref{qp}. Evidently, these conditions are fulfilled for both types of quark matter under different $m_{s0}$ within the quasiparticle model.

\begin{figure*}[t]
	\subfloat[\label{fig1a}]{\includegraphics{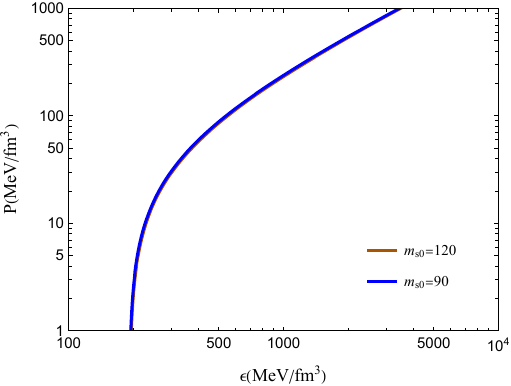}}
	\subfloat[\label{fig1b}]{\includegraphics{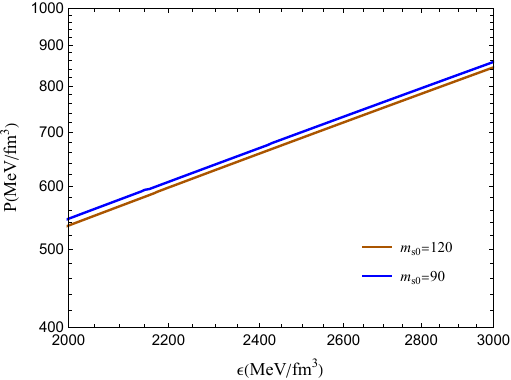}}
	\caption{\label{figqm} Equation of state for quark matter described by the quasiparticle model. The color of the lines indicates different parameter choices, with $m_{s0}=120$ shown in dark orange and $m_{s0}=90$ in blue. Panel b is an enlargement of a certain interval from panel a, in order to more clearly show the differences.}
\end{figure*}

The resulting EOSs with are presented in Fig~.\ref{figqm}, showing only tiny differences between them.

\subsection{Extended MIT bag model}
In the extended MIT bag model, medium effects are taken into account in the framework of MIT bag model \cite{schertlerQuarkPhasesNeutron2000}. Effective masses generated by the interaction of particles with the system is one of the most important medium effects \cite{schertlerMediumEffectsStrange1997}. Thus the cold strange quark matter can be treated as a system composed of non-interacting quarks with effective masses \cite{schertlerMediumEffectsStrange1997,schertlerStrangeMatterEffective1997}.

The effective quark mass of this extended MIT bag model reads \cite{schertlerMediumEffectsStrange1997,schertlerStrangeMatterEffective1997,blaizotCollectiveFermionicExcitations1993}, 
\begin{eqnarray}
	m_f=\frac{m_{f0}}{2}+\sqrt{\frac{m_{f0}^2}{4}+\frac{g^2\mu_f^2}{6\pi^2}},\label{eq11}
\end{eqnarray}
where $m_{f_0}$ is the current quark mass (here we take $m_{u0}=5.5 ~\mathrm{MeV}, m_{d0}=5.5~\mathrm{MeV}, ~\mathrm{and}~ m_{s0}=95~\mathrm{MeV}$ from Ref.~\cite{zhangQuarkMatterQuark2021}), $\mu_f$
represents the chemical potential of the different flavors of quark, and $g$ is the strong coupling constant which is treated as a free parameter in this work. 

The total thermodynamic potential density for strange quark matter can be written as
\begin{eqnarray}
	\Omega=\sum_i[\Omega_i+B_i(\mu_i)]+B,
\end{eqnarray}
where $B_i(\mu_i)$ are the additional terms necessary to maintain thermodynamic self-consistency, $B$ is a free parameter representing the vacuum pressure in the original MIT bag model \cite{patraTemperatureBaryonchemicalpotentialdependentBag1996}. And $\Omega_i$ denotes the thermodynamic potential density contributions from different flavors of quark ($u,~d, ~\mathrm{and}~ s$) and leptons ($e ~\mathrm{and}~ \mu$). The analytic expression for $\Omega_i$ is given by
\begin{eqnarray}
	\begin{aligned}\Omega_{i}&=-\frac{g_{i}}{48\pi^{2}}\left[\mu_{i}\sqrt{\mu_{i}^{2}-m_{i}^{2}}(2\mu_{i}^{2}-5m_{i}^{2})\right]\\&+3m_{i}^{4}\ln\frac{\mu_{i}+\sqrt{\mu_{i}^{2}-m_{i}^{2}}}{m_{i}^{2}}\biggr],\end{aligned}
\end{eqnarray}
where $g_i$ is the degeneracy factor with $g_i=6$ for quarks and $g_i=2$ for leptons. The $\mu$-dependent term $B_i(\mu_i)$ is
determined by using the integration formula as \cite{schertlerMediumEffectsStrange1997,zhangQuarkMatterQuark2021}
\begin{eqnarray}
	B_i(\mu_i)=-\int_{m_i(\mu_i)}^{\mu_i}\frac{\partial\Omega_i}{\partial m_i}\frac{\partial m_i(\mu')}{\partial\mu'}\mathrm{d}\mu'.
\end{eqnarray}
The total energy density $\epsilon$ and the pressure $P$ are expressed as
\begin{eqnarray}
	&&
	P=-\sum_{i}(\Omega_{i}+B_{i}(\mu_{i}))-B,\\
	&&
	\epsilon=\sum_{i}(\mu_{i}\rho_{i})-P.
\end{eqnarray}

\begin{figure}
	\includegraphics[width=0.5\textwidth]{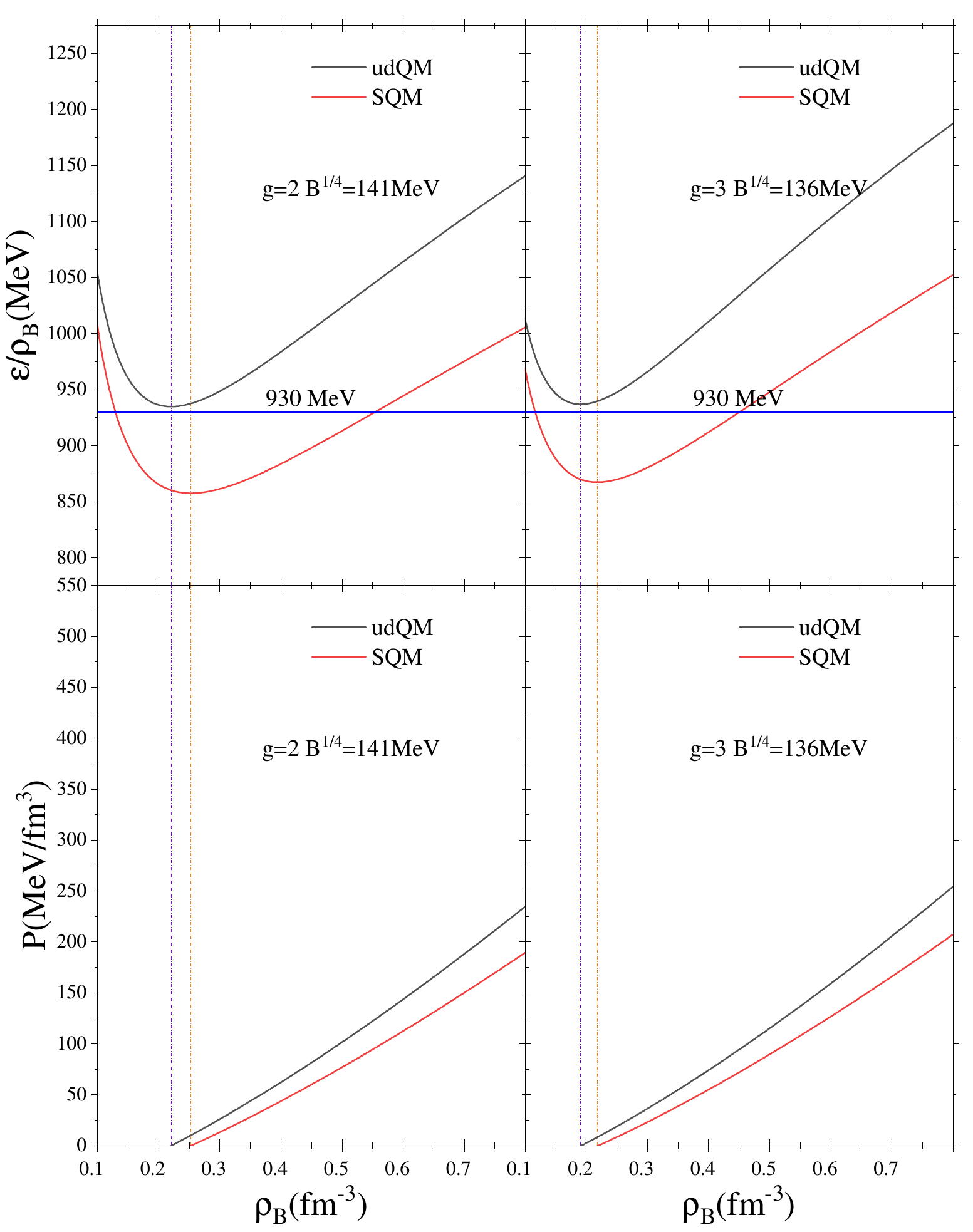}
	\caption{\label{ebm}The energy per baryon $\varepsilon/\rho_B$ and pressure $P$ as functions of baryon number density ($\rho_B$) for SQM (red line) and udQM (black line) described by the quasiparticle model at zero temperature. The parameter set is  $g=2,~B^{1/4}=141~\mathrm{MeV}$ for left panels and  $g=3,~B^{1/4}=136~\mathrm{MeV}$ for right panels. }
\end{figure}

Similar to the quasiparticle model, $\varepsilon/\rho_B$ and $P$ as functions of $\rho_B$ are presented first, and check whether the previously mentioned conditions are satisfied. As can be seen from Fig.~\ref{ebm}, the absolute stability conditions are fulfilled for both types of quark matter, and $\rho_B$ at zero-pressure point coincides with the $\rho_B$ at the point of minimum energy per baryon. With these verifications completed, one can turn to the EOS calculations.

\begin{figure}[h]
	\includegraphics{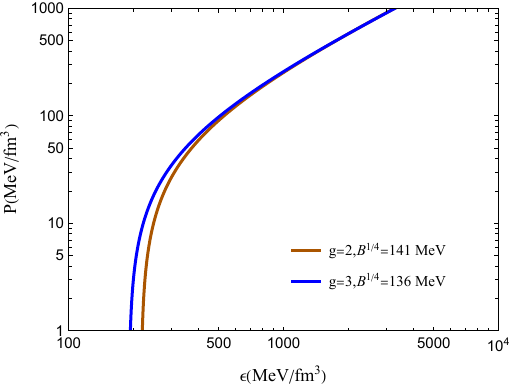}
	\caption{\label{figebm} Equations of state for quark matter described by the extended MIT bag model. The color of the lines indicates different parameter choices, with $g=2,~B^{1/4}=141~\mathrm{MeV}$ shown in dark orange and $g=3,~B^{1/4}=136~\mathrm{MeV}$ in blue.}
\end{figure}

The resulting EOSs are presented in Fig~.\ref{figebm}.
\section{Dark matter\label{sec3}}
As mentioned in the introduction, the dark matter considered in this work is asymmetric. Asymmetric dark matter may be composed of weakly-interacting massive particle \cite{petrakiREVIEWASYMMETRICDARK2013}. We assume these weakly-interacting massive particles are fermions with a mass of 100 GeV (as in Ref.~\cite{mukhopadhyayQuarkStarsAdmixed2016}). First we take the dark matter as a free Fermi gas, its energy density, pressure and number density can be expressed as,
\begin{eqnarray}
	\epsilon&=&\frac{1}{\pi^{2}}\int_{0}^{k_{F}}k^{2}\sqrt{m_{f}^{2}+k^{2}}dk\nonumber\\
	&=&\frac{m_{f}^{4}}{8\pi^{2}}\left[(2z^{3}+z)\sqrt{1+z^{2}}-\sinh^{-1}(z)\right]\\
	P&=&\frac{1}{3\pi^{2}}\int_{0}^{k_{F}}\frac{k^{4}}{\sqrt{m_{f}^{2}+k^{2}}}dk\nonumber\\
	&=&\frac{m_f^4}{24\pi^2}\left[(2z^3-3z)\sqrt{1+z^2}+3\mathrm{sinh}^{-1}(z)\right]\\
	n&=&\frac{k_F^3}{3\pi^2}\equiv\frac{m_f^3 z^3}{3\pi^2},
\end{eqnarray}
where $z=k_F/m_f$ is the dimensionless Fermi momentum, $k_F$ and $m_f$ are the Fermi momentum and masses of dark fermions.
Next we include interactions between dark fermions. To proceed, we consider the simplest two-body repulsive interactions \cite{narainCompactStarsMade2006}. In the lowest order approximation, the energy density of self-interaction is proportional to $n^2$ \cite{narainCompactStarsMade2006,agnihotriBosonStarsRepulsive2009}. The resulting equation of state is given by \cite{mukhopadhyayQuarkStarsAdmixed2016}
\begin{eqnarray}
	\epsilon&=&\frac{m_{f}^{4}}{8\pi^{2}}\left[(2z^{3}+z)\sqrt{1+z^{2}}-\sinh^{-1}(z)\right]\nonumber\\
	&&
	+\left[\left(\frac{m_f^3z^{3}}{3\pi^{2}}\right)^{2}\frac{\eta^{2}}{m_f^2}\right] \\
	P&=&\frac{m_{f}^{4}}{24\pi^{2}}\left[(2z^{3}-3z)\sqrt{1+z^{2}}+3\sinh^{-1}(z)\right]\nonumber\\
	&&
	+\left[\left(\frac{m_f^3z^3}{3\pi^2}\right)^2\frac{\eta^{2}}{m_f^2}\right],
\end{eqnarray}
where $\eta=m_f/m_I$ is the dimensionless interaction strength parameter and $m_I$ is the interaction mass scale \cite{narainCompactStarsMade2006,marianiConstrainingSelfinteractingFermionic2023,wangPossibleMaximumMass2019,mukhopadhyayQuarkStarsAdmixed2016}.

\begin{figure}[t]
	\includegraphics{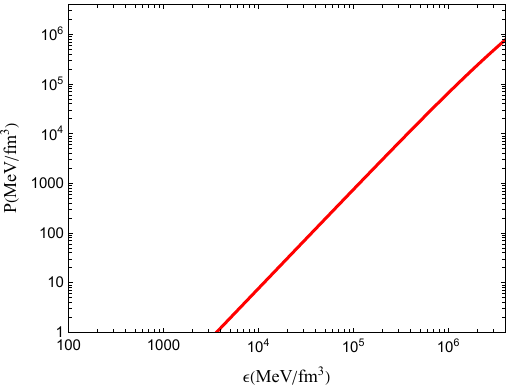}
	\caption{\label{figdm100} Equation of state for fermionic dark matter with strong self-interaction.}
\end{figure}

The typical interaction mass scale of strong interaction is $m_I\sim100~\mathrm{MeV}$. In this case, dark matter fermions with a mass of 100 GeV can have strong self-interactions with a strength of $\eta\sim10^3$. Thus, here we focus on the situation of strongly self-interacting dark matter with $m_I=100~\mathrm{MeV}$ and $\eta=10^3$. The resulting EOS in this case is presented in Fig~.\ref{figdm100}.

\section{Two-fluid TOV \textbf{equations} and tidal deformability\label{sec4}}
In order to obtain the properties of a strange quark star admixed with dark matter, the two-fluids TOV equations are needed. In a two-fluid system, each independent fluid has its own hydrodynamic equilibrium condition, and there are only gravitational interactions between the fluids. The two-fluid TOV equations that we use here are \cite{sandinEffectsMirrorDark2009,tolosDarkCompactPlanets2015,mukhopadhyayQuarkStarsAdmixed2016,hippertDarkMatterRegular2023,marianiConstrainingSelfinteractingFermionic2023,hongMixedDarkMatter2024}
\begin{eqnarray}
	\frac{dp_{1}}{dr}&=&-\frac{G}{r^{2}}\left[M(r)+4\pi r^3(p_1(r)+p_2(r))\right]\nonumber \\
	&&
	\times\left[\epsilon_{1}(r)+p_{1}(r)\right]\left[1-2G\frac{M(r)}{r}\right]^{-1} \\
	\frac{dp_{2}}{dr}&=&-\frac{G}{r^{2}}\left[M(r)+4\pi r^3(p_1(r)+p_2(r))\right] \nonumber\\
	&&
	\times\left[\epsilon_{2}(r)+p_{2}(r)\right]\left[1-2G\frac{M(r)}{r}\right]^{-1} \\
	\frac{dM_1}{dr}&=&4\pi r^2\epsilon_1(r)  \\
	\frac{dM_2}{dr}&=&4\pi r^2\epsilon_2(r)  \\
	M(r)&=&M_1(r)+M_2(r).  
\end{eqnarray}
To solve these equations, we need the boundary conditions $M_1(0) = 0$ and $M_2(0) = 0$, as well as the values of central pressures $p_1(0)$ and $p_2(0)$ . Then the functions of $p_1(r),~p_2(r)~\mathrm{and}~M(r)$ can be solved. The pressures of both fluids drop to zero at radial coordinate points $R_1$ and $R_2$ respectively, i.e., $p_1(R_1)=p_2(R_2)=0$. The larger value among $R_1$ and $R_2$ is defined as the overall admixed star radius $R_{out}$. Finally, the total mass of the admixed star can be obtained by substituting $R_{out}$ into $M(r)$.

In addition to mass and radius, tidal deformability is also a property we are interested in. In order to obtain the tidal deformability of the star, we need to solve the following equation along with the TOV equation simultaneously \cite{croninRotatingDarkMatter2023,hindererTidalLoveNumbers2008}:
\begin{eqnarray}
	&&
	r\frac{dy(r)}{dr}+y^2(r)+y(r)F(r)+r^2Q(r)=0,\label{eq27}\\
	F(r)&=&\frac{r-4\pi Gr^3\sum_{i}(\epsilon_i-p_i)}{r-2GM}\\
	Q(r)&=&\frac{4\pi Gr}{r-2GM}\nonumber\\
	&&\times\left\{\sum_{i}\left[5\epsilon_i+9p_i+\frac{\partial\epsilon_i}{\partial p_i}\left(\epsilon_i+p_i\right)\right]-\frac{6}{4\pi Gr^{2}}\right\}\nonumber\\
	&&-\left[\frac{8\pi Gr^{3}\sum_{i}p_i+2GM}{r(r-2GM)}\right]^{2}\label{eq29}.
\end{eqnarray}
The sum over $i$ in functions $F(r)$ and $Q(r)$ implies considering multiple fluids. Here $y(r)$ is the logarithmic derivative for metric perturbation \cite{takatsyCommentTidalLove2020,damourRelativisticTidalProperties2009}.

There is the squared adiabatic speed of sound $\frac{d\epsilon}{dp}$ in Eq.~(\ref{eq27}) for $y(r)$. It should be noted that according to models of strange quark stars, there is a finite energy density at their surface. Such situation similar to a first-order phase transition at constant pressure leads to $c_s^2\equiv\frac{dp}{d\epsilon}$, while $\frac{d\epsilon}{dp}$ is discontinuous. To address this, the expression needs to be modified as follow \cite{postnikovTidalLoveNumbers2010,takatsyCommentTidalLove2020}
\begin{eqnarray}
	\frac{d\epsilon}{dp}=\frac{1}{c_{s}^{2}}=\frac{d\epsilon}{dp}\Bigg|_{p\neq p_{d}}+\Delta\epsilon\delta(p-p_{d}).\label{eq30}
\end{eqnarray} 
The $\Delta\epsilon\delta(p-p_{d})$ term results in an extra term in the solution of $y(r)$, which is written as \cite{takatsyCommentTidalLove2020}:
\begin{eqnarray}
	y(r_d^+)-y(r_d^-)=-\frac{4\pi r_d^3\Delta\epsilon}{m(r_d)+4\pi r_d^3p(r_d)}.
\end{eqnarray}
Here $r_d$ is the position where $\frac{d\epsilon}{dp}$ is discontinuous, and $r_d^\pm$ represents two points infinitesimally distant from $r_d$ in opposite directions. For a admixed quark star, $r_d$ here represents the radius of the portion containing strange quark matter ($R_{qm}$). And $\Delta\epsilon=\epsilon(R_{qm}^-)$ is the energy density just inside this strange quark matter region.

After obtaining the solution at surface of the star 
\begin{eqnarray}
	y_R=y(R_{out})+extra~term,\nonumber
\end{eqnarray}
and substitute it into the expression for the $l = 2$ tidal Love number,
\begin{eqnarray}
		k_{2}& =&\frac85\beta^5(1-2\beta)^2[2-y_R+2\beta(y_R-1)]\nonumber \\
		&&\times\{2\beta[6-3y_R+3\beta(5y_R-8)]\nonumber \\
		&&+4\beta^2[13-11y_R+\beta(3y_R-2)+2\beta^2(1+y_R)]\nonumber \\
		&&+3(1-2\beta)^2[2-y_R+2\beta(y_R-1)]\ln(1-2\beta)\}^{-1},\nonumber\\
		&&
\end{eqnarray}
where $\beta=GM(R_{out})/R_{out}$. Finally, the tidal deformability of the star can be calculated through the formula, which reads:
\begin{eqnarray}
	\Lambda=\frac{2}{3}k_2\beta^{-5}.
\end{eqnarray}
\section{Properties of strange quark stars with dark matter\label{sec5}}
In this section, we use the two quark models introduced above to investigate the properties of strange quark stars with fermionic dark matter inside.

To study the influence of dark matter on the properties of strange quark stars, observational data is essential for comparison. A compact object has recently been discovered at the center of the supernova remnant HESS J1731-347 \cite{doroshenkoStrangelyLightNeutron2022}. It is speculated to be either the lightest known neutron star or a strange quark star. The gravitational wave event GW170817 \cite{PhysRevLett.121.161101,PhysRevLett.119.161101} is also an important data source. Existing analyses based on this event have yielded a range of masses and radii, as well as an interval for tidal deformability of a $1.4M_{\odot}$ compact star $\Lambda_{1.4}=190_{-120}^{+390}$. In addition to the aforementioned data, in this work we also employ data concerning PSR J0348+0432 ($2.01\pm0.04$ solar mass $M_\odot$) \cite{antoniadisMassivePulsarCompact2013} and PSR J0030+0451 \cite{millerPSRJ003004512019, rileyNICERViewPSR2019}. See the figures drawn later in this section for details.
\subsection{Quasiparticle model}
As mentioned earlier, the current strange quark mass $m_{s0}$ in the quasiparticle model is treated as a free parameter. Here we compare the numerical results with $m_{s0}$ set to 120 MeV and 90 MeV respectively.

First, we study the relationship between the mass of admixed quark stars and their central energy density. The central energy density is divided into that of quark matter ($\epsilon_{qm0}$) and dark matter ($\epsilon_{dm0}$). And contour lines of stellar mass can be drawn on the $\epsilon_{dm0}$-$\epsilon_{qm0}$ plane.
\begin{figure*}[ht]
	\centering
	\includegraphics[width=\textwidth]{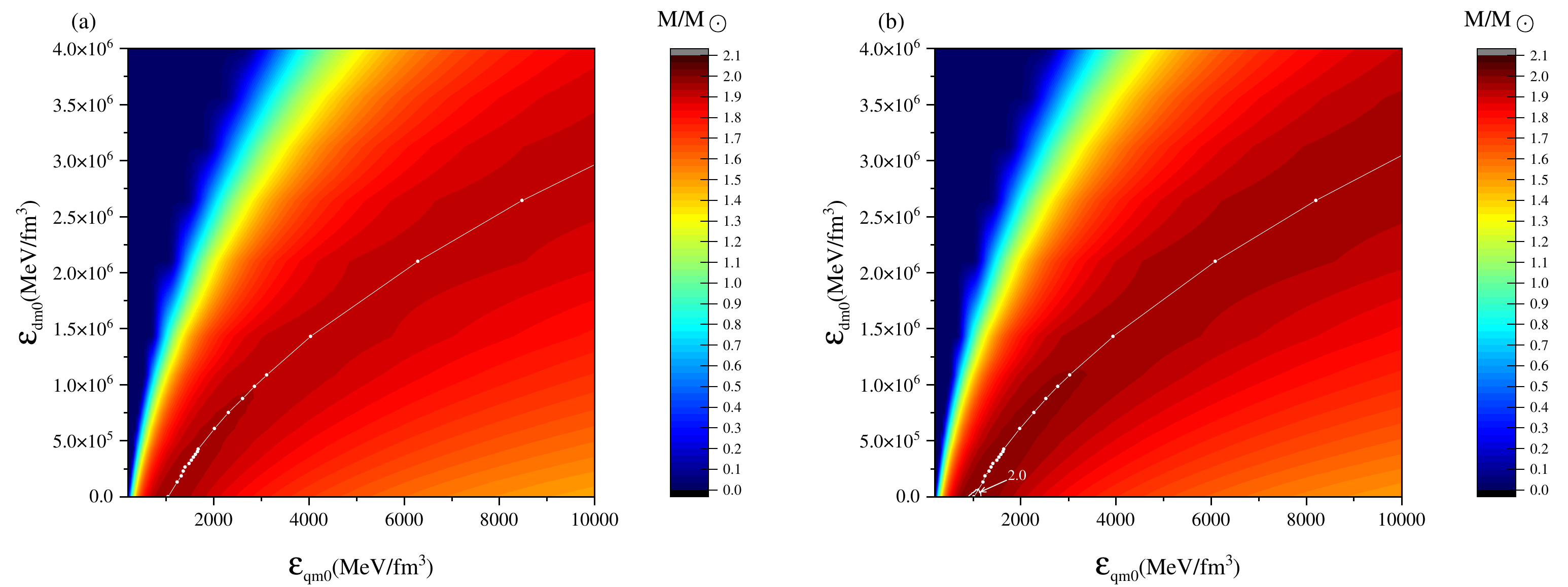}
	\caption{\label{fig1} Contour plot with $\epsilon_{qm0}$, $\epsilon_{dm0}$ and $M/M_{\odot}$ as the x, y, and z axes, respectively. The current strange quark mass $m_{s0}$ is 120 MeV for panel (a) and 90 MeV for panel (b). The white line with scattering points represents the maximum mass of quark stars admixed with dark matter. Such stars in the region to the right of this line are unstable. The region where the maximum mass exceeds $2M_{\odot}$ is delineated by a white curve.}
\end{figure*}

Then the dependence of the total stellar mass (in solar mass $M_{\odot}$) on the central energy densities of the two fluids is presented in Fig.~\ref{fig1}. The region to the right of the white scattered line represents unstable stars and is not considered. It is evident that when $\epsilon_{qm0}$ is fixed, increasing $\epsilon_{dm0}$ leads to a decrease in the total stellar mass. Conversely, when $\epsilon_{dm0}$ is fixed, changing $\epsilon_{qm0}$ produces the opposite effect. It is also noted that the impact on mass due to changes in $\epsilon_{dm0}$ is far less significant than that caused by changes in $\epsilon_{qm0}$.
\begin{figure*}[t]
	\centering
	\subfloat[\label{fig2a}]{\includegraphics{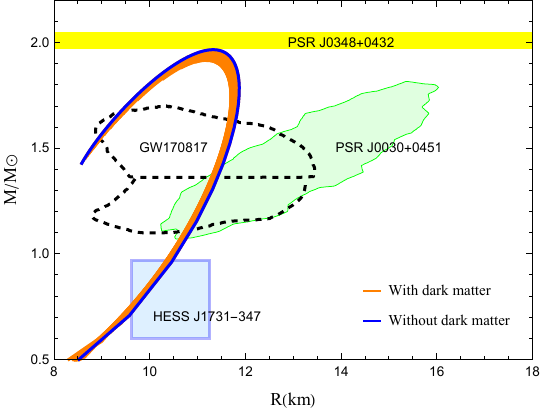}}
	\subfloat[\label{fig2b}]{\includegraphics{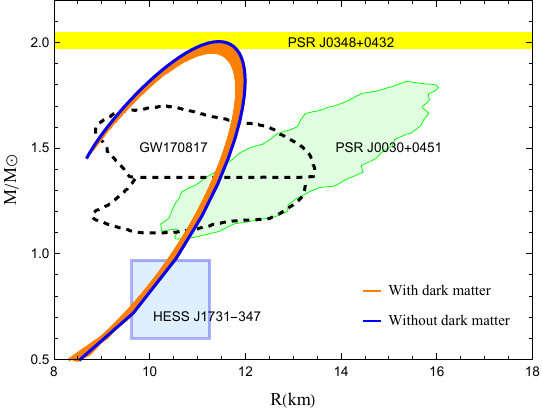}}
	\caption{Mass-radius ($M-R$) relation of strange quark star admixed with dark matter. The current strange quark mass $m_{s0}$ is 120 MeV for panel (a) and 90 MeV for panel (b). The color of the solid line represents the strange quark star admixed with (orange) or not admixed with (blue) dark matter. The light green region represents the observational data of PSR J0030+0451 after Bayesian analysis \cite{millerPSRJ003004512019}, while the light blue region is from the observational data of the central compact object within the supernova remnant HESS J1731-347 \cite{doroshenkoStrangelyLightNeutron2022}. The area enclosed by the black dashed line comes from the gravitational wave data analysis of GW170817 \cite{PhysRevLett.121.161101,PhysRevLett.119.161101} and the yellow band corresponds to the mass measurement of PSR J0348+0432 \cite{antoniadisMassivePulsarCompact2013}. \label{fig2}}
\end{figure*}

Next, we present the relation between mass and radius in Fig.~\ref{fig2}. Multi-messenger astronomical observation data are presented in the figure as regions/bands of different colors. The first and most obvious point is that the $M-R$ curves of admixed quark stars are consistent with most of the observational data shown in Fig.~\ref{fig2} . However, some of the maximum masses cannot reach the data of PSR J0348+0432. This is partly due to the choice of parameters in the quark model, where a larger $m_{s0}$ would reduce the maximum mass of the star. On the other hand, the addition of dark matter also reduces the maximum mass of admixed quark stars. And the $M-R$ curve of quark stars without dark matter almost envelops that of those with dark matter.

\begin{figure*}[t]
	\centering
	\subfloat[\label{fig3a}]{\includegraphics{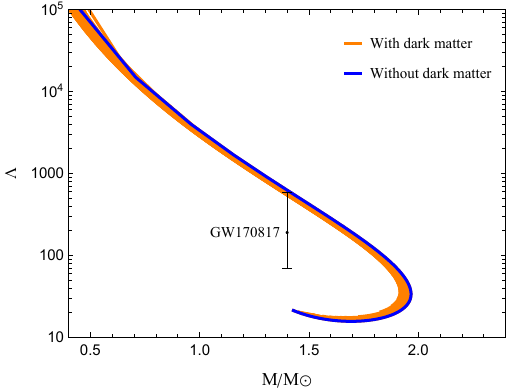}}
	\subfloat[\label{fig3b}]{\includegraphics{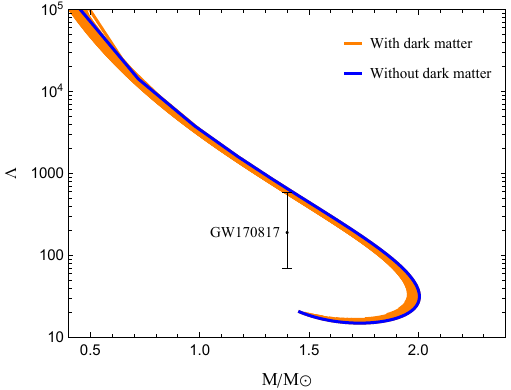}}
	\caption{Relation between tidal deformability and the mass ($\Lambda-M$) of admixed quark star. The current strange quark mass $m_{s0}$ is 120 MeV for panel (a) and 90 MeV for panel (b). The black error bar represents the range of tidal deformability for GW170817 at 1.4$M_{\odot}$ \cite{PhysRevLett.121.161101}. The color of the solid line represents the strange quark star admixed with (orange) or not admixed with (blue) dark matter. \label{fig3}}
\end{figure*}

To further investigate the gravitational effects of adding dark matter to strange quark stars, the relationship between tidal deformability ($\Lambda$) and mass is presented in Fig.~\ref{fig3}. It can be seen that for 1.4$M_{\odot}$ admixed quark stars, increased dark matter and smaller $m_{s0}$ reduce tidal deformability, aligning partially with GW170817 data. And corresponding to the position of 1.4$M_{\odot}$ on the $M-R$ curve, it can also be observed that the decrease in tidal deformability is accompanied by a reduction in radius.

\subsection{Extended MIT bag model}
\begin{figure*}[ht]
	\centering
	\includegraphics[width=\textwidth]{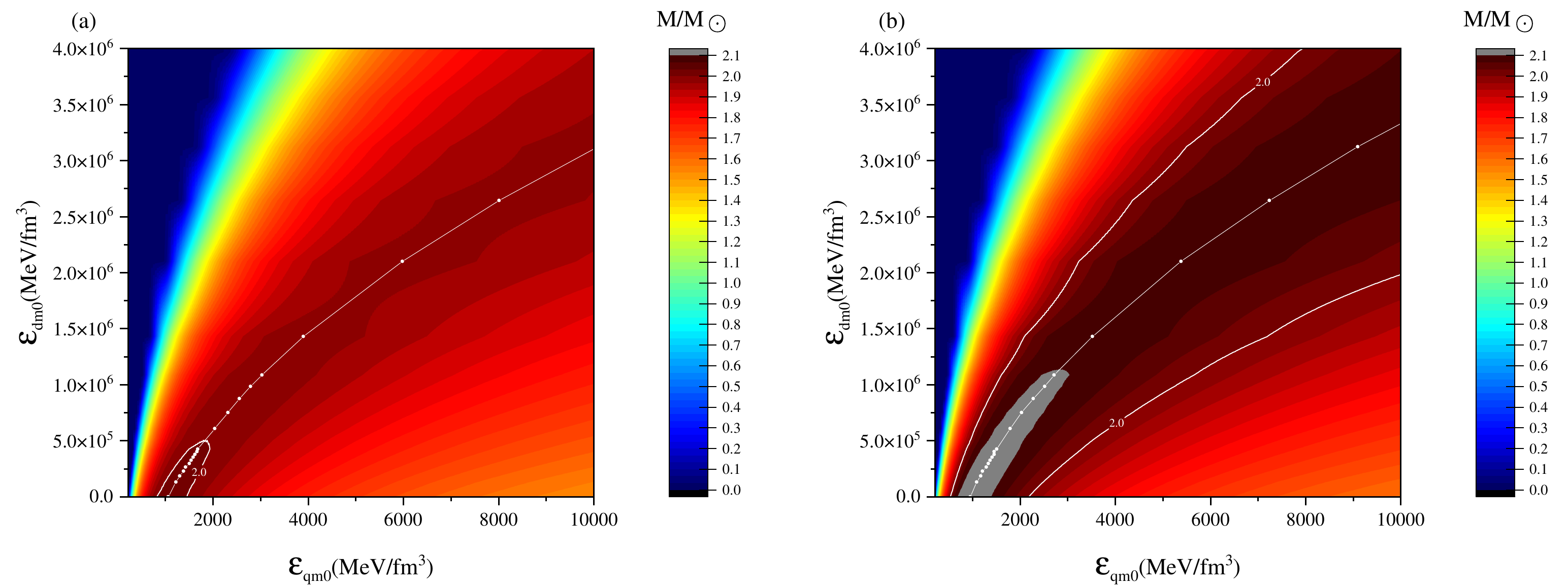}
	\caption{\label{fig4} Contour plot with $\epsilon_{qm0}$, $\epsilon_{dm0}$ and $M/M_{\odot}$ as the x, y, and z axes, respectively. The parameter set is $g=2,~B^{1/4}=141~\mathrm{MeV}$ for panel (a) and $g=3,~B^{1/4}=136~\mathrm{MeV}$ for panel (b). The white line with scattering points represents the maximum mass of quark stars admixed with dark matter. Such stars in the region to the right of this line are unstable. The region where the maximum mass exceeds $2M_{\odot}$ is delineated by a white curve.}
\end{figure*}

\begin{figure*}[!t]
	\centering
	\subfloat[\label{fig5a}]{\includegraphics{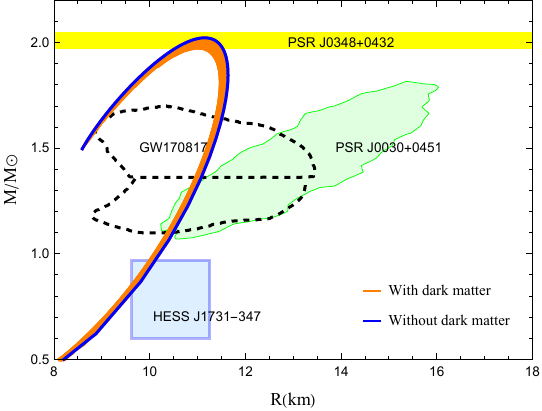}}
	\subfloat[\label{fig5b}]{\includegraphics{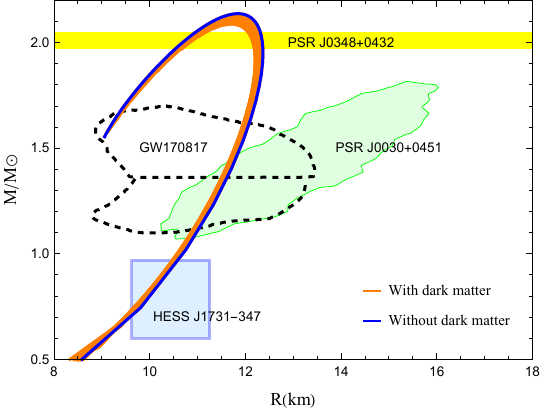}}
	\caption{Mass-radius ($M-R$) relation of strange quark star admixed with dark matter. The parameter set is $g=2,~B^{1/4}=141~\mathrm{MeV}$ for panel (a) and $g=3,~B^{1/4}=136~\mathrm{MeV}$ for panel (b). The color of the solid line represents the strange quark star admixed with (orange) or not admixed with (blue) dark matter. The light green region represents the observational data of PSR J0030+0451 after Bayesian analysis \cite{millerPSRJ003004512019}, while the light blue region is from the observational data of the central compact object within the supernova remnant HESS J1731-347 \cite{doroshenkoStrangelyLightNeutron2022}. The area enclosed by the black dashed line comes from the gravitational wave data analysis of GW170817 \cite{PhysRevLett.121.161101,PhysRevLett.119.161101} and the yellow band corresponds to the mass measurement of PSR J0348+0432 \cite{antoniadisMassivePulsarCompact2013}. \label{fig5}}
\end{figure*}

\begin{figure*}[!t]
	\centering
	\subfloat[\label{fig6a}]{\includegraphics{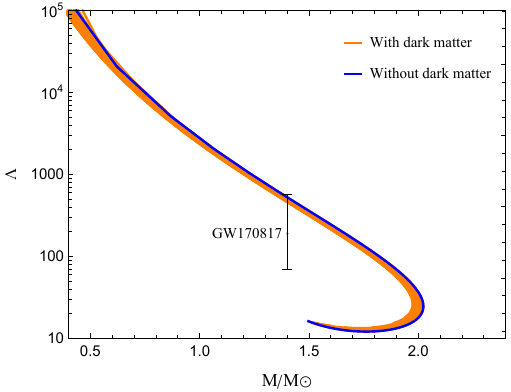}}
	\subfloat[\label{fig6b}]{\includegraphics{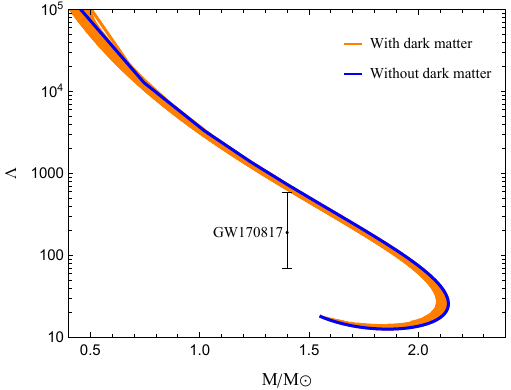}}
	\caption{Relation between tidal deformability and the mass ($\Lambda-M$) of admixed quark star. The parameter set is $g=2,~B^{1/4}=141~\mathrm{MeV}$ for panel (a) and $g=3,~B^{1/4}=136~\mathrm{MeV}$ for panel (b). The black error bar represents the range of tidal deformability for GW170817 at 1.4$M_{\odot}$ \cite{PhysRevLett.121.161101}. The color of the solid line represents the strange quark star admixed with (orange) or not admixed with (blue) dark matter. \label{fig6}}
\end{figure*}

When describing quark matter using this model, we employ two sets of parameters: $g=2,~B^{1/4}=141~\mathrm{MeV}$ and $g=3,~B^{1/4}=136~\mathrm{MeV}$. Similar to the previous subsection, here we also present three figures illustrating the following relationships: $\epsilon_{dm0}-\epsilon_{qm0}-M$ (Fig.~\ref{fig4}), $M-R$ (Fig.~\ref{fig5}) and $\Lambda-M$ (Fig.~\ref{fig6}).

In Fig.~\ref{fig4}, we can observe that the trend of maximum mass variation with central energy density ($\epsilon_{qm0}~\mathrm{and}~\epsilon_{dm0}$) is consistent with that of the quasiparticle model. Subsequent analysis focuses on the $M-R$ plot (Fig.~\ref{fig5}) and $\Lambda-M$ plot (Fig.~\ref{fig6}). The pattern of how dark matter content affects the two curves is similar to that in the quasi-particle model. Both the maximum mass and $\Lambda_{1.4}$ will slightly decrease as the amount of dark matter increases. However, comparatively, changes in the parameter set have a much greater impact on stellar properties than changes in the amount of dark matter.
It can be found in Fig.~\ref{fig6} that once $\Lambda_{1.4}$ (without dark matter) exceeds a certain threshold, increasing dark matter content fails to reduce $\Lambda_{1.4}$ to the observational data range.

\section{Summary and discussion\label{sec6}}
In this work, we have investigated the effects of dark matter on strange quark stars. The strange quark matter is described by the quasiparticle model and the extended MIT bag model, respectively. Dark matter is assumed to be self-interacting Fermi gas, and interacts with quark matter only through gravitational interactions. The two-fluid TOV equations are solved to study the properties of strange quark stars admixed with dark matter.

Both the quasiparticle model and the extended MIT bag model employ the quasiparticle concept and account for medium effects. The impact of fermionic asymmetric dark matter on strange quark stars described by these models may follow a similar pattern. For a fixed $\epsilon_{qm0}$, the mass of stable admixed quark star decreases as the $\epsilon_{dm0}$ increases. To maintain a constant mass while increasing the $\epsilon_{dm0}$, a higher $\epsilon_{qm0}$ is required. Similarly, the increase in $\epsilon_{dm0}$ drives stable admixed quark stars towards greater $\epsilon_{qm0}$ to attain maximum mass. Concurrently, the maximum mass gradually decreases with increasing central energy density of dark matter. Besides the maximum mass, $\Lambda_{1.4}$ also exhibits this pattern, i.e., The increase in the central energy density of dark matter causes $\Lambda_{1.4}$ to decrease.

Evidently, most of our numerical results for $M-R$ and $\Lambda-M$ relations are consistent with observational data. It had been expected that the addition and increase of fermionic asymmetric dark matter would significantly alter the calculated $M-R$ curve and $\Lambda-M$ curve. However, our calculations in this work show that its impact appears to be less significant. Changes in certain parameters of the quark model used in this work appear to have an even greater impact on these curves. Consequently, there is still significant room for future investigation into how different types of dark matter and their parameter variations affect strange quark stars described by various quark models. 
It is important for the study of properties of dark matter and strange quark stars.
\bibliography{qmdm}

\end{document}